# Reductive Contact and Dipolar Interface Engineering Enable Stable Flexible CsSnI$_3$ Nanowire Photodetectors

*Letian Dai[a,b,*], Wanru Chen[c], Quanming Geng[a], Ying Xu[a], Guowu Zhou[d], Nuo Chen[e], Xiongjie Li[f,*]*

[a] Wuhan National Laboratory for Optoelectronics, Huazhong University of Science and Technology, 1037 Luoyu Road, Wuhan 430074, Hubei, P. R. China

[b] Optics Valley Laboratory, Hubei 430074, P. R. China

[c] Wuhan Huaming Renewables Co., Ltd., 122 Zhongbei Road, Wuhan 430077, Hubei, P.R. China

[d] Wuhan Haisipulin Technology Development Co., Ltd, 897 Gexin Road, Wuhan, 430023, Hubei, P.R. China

[e] Department of Electrical Engineering, Karlsruhe Institute of Technology, Karlsruhe 76131, Germany

[f] Key Laboratory of Catalysis and Energy Materials Chemistry of Ministry of Education & Hubei Key Laboratory of Catalysis and Materials Science, Key Laboratory of Analytical Chemistry of the State Ethnic Affairs Commission, Hubei R&D Center of Hyperbranched Polymers Synthesis and Applications, South-Central Minzu University, Wuhan, Hubei, 430074, P.R. China

*Email: ldai@hust.edu.cn, xiongjieli@mail.scuec.edu.cn.

**Abstract**

Lead-free tin-based halide perovskites are attractive for flexible and environmentally benign optoelectronics, but their application is limited by the rapid oxidation of $Sn^{2+}$ to $Sn^{4+}$ and poor operational stability. Here, we report a flexible $CsSnI_3$ nanowire photodetector that achieves both high near-infrared photoresponse and long-term stability through synergistic aluminium-substrate contact engineering and dipolar interface modification. A 0.2 mm anodized aluminium foil serves as the flexible substrate, where localized laser ablation exposes metallic aluminium regions that act as reductive sites, effectively suppressing $Sn^{2+}$ oxidation during nanowire growth. Simultaneously, a polar interlayer of 3-fluoro-2-nitroanisole is introduced to improve energy-level alignment, suppress interfacial deprotonation, and enhance charge extraction. The resulting device exhibits a responsivity of 0.39 A W$^{-1}$, a specific detectivity of $1.38 \times 10^{13}$ Jones, and a wide linear dynamic range of 156 dB under 850 nm illumination. Moreover, the device retains over 85% of its initial photocurrent after 60 days in ambient air and maintains 94% of its initial photocurrent after 1000 bending cycles. This work establishes an effective strategy for stabilizing Sn-based perovskites toward high-performance flexible optoelectronic devices.

**Introduction**

In recent years, flexible photodetectors have become a major focus driven by the rapid evolution of flexible electronics, where flexible displays, wearable health monitors, smart textiles, and conformable sensors are emerging as core components of next-generation consumer and medical technologies [1–5]. Over the past decade, perovskite-based photodetectors have undergone significant development, demonstrating exceptional properties such as high photoresponsivity[6,7], fast response speed[8,9], and tunable spectral sensitivity[10,11], thereby attracting considerable attention from both academic and industrial communities. However, the vast majority of these perovskite photodetectors rely on lead halide perovskite materials as their active layers. Given the close and prolonged contact between wearable devices and human skin, the lead component in these materials poses non-negligible health and safety risks for wearable applications.[12,13] Notably, for wearable and close-contact applications, stringent regulations on heavy metal content (e.g., Restriction of Hazardous Substances Directive, RoHS[14]) impose strict limits on lead (Pb) concentrations—often below 1000 ppm—due to potential skin absorption or accidental ingestion risks during long-term use. This regulatory and safety-driven demand further emphasizes the urgency of replacing lead-based perovskites with non-toxic alternatives in flexible optoelectronic devices.

Given the environmental advantages and promising optoelectronic properties of tin-based perovskites, $CsSnI_3$ has received particular attention[15–17]. Its suitable (~1.3 eV)[18,19], high absorption coefficients[20], and high carrier mobility[21,22] make it one of the most representative inorganic tin-based candidates.[23–28] However, $CsSnI_3$ intrinsically exhibits p-type doping, originating from the easy oxidation of $Sn^{2+}$ to $Sn^{4+}$ during synthesis and ambient exposure, which induces the formation of tin vacancies ($V_{Sn}$).[29–31] In such p-type systems, photogenerated carriers are dominated by holes; therefore, selecting an appropriate hole transport layer (HTL) is crucial to improve hole extraction efficiency and suppress interfacial recombination.

Prior to this work，flexible perovskite photodetectors were primarily based on Pb-based perovskites,[32–34] and the deployment of tin-based perovskites in flexible optoelectronic devices has remained highly limited. This is mainly due to the inherent instability of $Sn^{2+}$, which is highly susceptible to oxidation into $Sn^{4+}$ in ambient conditions, resulting in deep defect formation and rapid performance degradation.[35–38]

To mitigate the defect accumulation arising from the oxidation of $Sn^{2+}$ to $Sn^{4+}$, a variety of stabilization strategies have been developed, including reducing agents[39,40], encapsulation[41,42], and modified processing atmospheres[43,44]. Karim *et al.* incorporated reducing additives into $FASnI_3$ precursor solutions to lower the $Sn^{4+}$ content[39], while Liu *et al.* employed polymer or $Al_2O_3$ encapsulation layers to block moisture and oxygen, thereby significantly improving the environmental stability of Sn-based films[42].

Despite these advances, several limitations persist. Additive engineering inevitably alters precursor chemistry, whereas encapsulation often relies on additional deposition steps that hinder compatibility with large-area or flexible device fabrication. More importantly, these approaches predominantly stabilize the bulk or surface of the perovskite layer, while insufficient attention has been directed toward the perovskite/substrate interface, where Sn oxidation is more likely to initiate.

Therefore, developing an interfacial strategy that can effectively suppress $Sn^{2+}$ oxidation while remaining compatible with simple, scalable, and flexible-device fabrication is essential for enhancing the stability of $CsSnI_3$-based optoelectronic devices. To address this challenge, we introduce a dual-interface engineering approach that integrates aluminium-substrate-induced reduction with dipolar molecular modification.

In this work, we report a tin-based perovskite flexible photodetector constructed from $CsSnI_3$ nanowires grown on a 0.2 mm-thick aluminium (Al) foil substrate. An insulating $Al_2O_3$ layer is first formed by anodization, followed by localized laser ablation to create well-defined growth windows that expose metallic Al, thereby establishing a reductive interface that suppresses $Sn^{2+}$ oxidation during $CsSnI_3$ crystallization. Following nanowire growth, a thin 3F-2NA molecular layer is introduced at the perovskite interface to generate an oriented dipole that modulates interfacial energy alignment and further enhances chemical stability. A compact $NiO_x$ film is subsequently deposited as the HTL, facilitating efficient hole extraction while blocking electron back-injection. Finally, a transparent poly(methyl methacrylate) (PMMA) coating is applied onto the indium tin oxide (ITO) top electrode to provide lightweight encapsulation suitable for flexible architectures.

Through the synergistic combination of substrate engineering, laser patterning, dipolar interface modification, HTL integration, and encapsulation, the resulting flexible $CsSnI_3$ nanowire photodetectors exhibit markedly enhanced optoelectronic performance and environmental stability. This work demonstrates a viable pathway toward high-performance, lead-free, and mechanically robust flexible optoelectronic devices.

**Results and discussion**

To begin with, we highlight the key conceptual advantages of our design strategy, as illustrated in Figure 1. A flexible aluminum substrate serves as both the mechanical support and a chemically active platform. Owing to the strong reducing nature of metallic Al, residual $Sn^{4+}$ species are spontaneously reduced to $Sn^{2+}$, effectively suppressing the undesired oxidation of $Sn^{2+}$ during and after film formation. Moreover, the use of vertically aligned $CsSnI_3$ nanowire arrays as the photoactive layer enables efficient light trapping and enhanced photon absorption, thereby improving the overall photoresponsivity of the device. To ensure stable operation in ambient conditions, a highly transparent and gas-impermeable PMMA overlayer is employed as the encapsulation layer, preventing moisture and oxygen ingress and further enhancing the environmental stability of the device. Overall, the aluminium substrate and PMMA encapsulation work synergistically to provide both chemical reduction and physical protection for the $CsSnI_3$ nanowire layer.

The 0.2 mm-thick 1060 aluminum foil (Figure S1a, Supporting Information) offers excellent mechanical strength, flexibility, and electrical conductivity, making it a suitable substrate for flexible optoelectronic devices. Four-point probe measurements show a sheet resistance of 0.63 mΩ/sq (bulk resistivity ~12.6 μΩ·cm), confirming its excellent conductivity. Mechanical tests, including uniaxial tension and three-point bending, yield a Young's modulus of ~68 GPa and a bending stiffness of ~4.6 × $10^{-3}$ N·m, consistent with reported values[45] and demonstrating outstanding flexibility. After anodization, the foil surface changes from a bright metallic appearance to a uniform grayish-white matte finish (Figure S1b). SEM images further reveal reduced surface roughness and the formation of a porous anodic aluminum oxide (AAO) layer with an average pore diameter of ~100 nm (Figure S2).

Energy-dispersive X-ray spectroscopy (EDX) analysis further confirmed the elemental composition and uniformity of the anodized aluminum foil, revealing distinct peaks corresponding to Al and O. Figure S3a shows the optical image of the anodized aluminum surface after laser patterning, where several circular laser-ablated regions are clearly observed. Figure S3b presents the corresponding SEM micrograph of a representative ablated spot, highlighting the well-defined circular boundary between the exposed metallic Al region and the surrounding anodic $Al_2O_3$ layer. As shown in the elemental mapping results (Figure S3c for Al and Figure S3d for O), oxygen is uniformly distributed across the oxidized area, consistent with the formation of a continuous anodic $Al_2O_3$ film. The corresponding EDX spectrum (Figure S3e) exhibits strong signals from Al and O without any detectable impurity peaks, further verifying the high purity and uniformity of the oxide layer.

Figure 2a illustrates the initial fabrication steps of the $CsSnI_3$ nanowire-based flexible photodetector, including AAO/Al substrate preparation, laser patterning, nanowire growth, interfacial modification, and $NiO_x$ HTL deposition. After anodization, the flexible aluminum substrate becomes electrically insulating due to the formation of an $Al_2O_3$ layer. To define conductive growth regions, localized laser ablation is employed to selectively remove this insulating layer, creating well-defined conductive windows that expose the underlying metallic Al. These laser-defined Al windows not only restore local electrical conductivity to a level comparable to pristine aluminum foil but also serve as effective reductive sites.

CsSnI$_3$ nanowires were grown using a strategy and methodology similar to those reported in our previous work.[24] We further investigated the wetting behavior of the CsSnI$_3$ precursor solution on different interfacial materials. As shown in Figure S4a and S4b, the CsSnI$_3$ precursor solution exhibits contact angles of 81.3° on metallic Al and 26.8° on AAO surface, reflecting a pronounced contrast in surface energy. The higher surface energy and polarity of the Al$_2$O$_3$ layer facilitate the spreading of the polar DMF solvent, whereas the larger contact angle on metallic Al results in confined precursor distribution within the laser-defined regions. After solvent evaporation, CsSnI$_3$ nanowires spontaneously form on both surfaces. SEM images (Figure 2b and Figure S5) reveal distinct growth behaviors: on metallic Al, CsSnI$_3$ adopts a predominantly vertical growth mode, yielding well-aligned nanowires exceeding 50 μm in length; in contrast, the anodized Al$_2$O$_3$ surface promotes lateral growth, forming microrods or microwires up to 5 μm in diameter and over 100 μm in length. This clear difference in morphology correlates directly with the contrasting wetting and precursor-spreading behaviors, demonstrating that substrate surface energy governs both nucleation and growth orientation.

Meanwhile, a droplet of CsSnI$_3$ precursor solution was deposited onto the laser-defined region, enabling direct contact between the precursor and the underlying metallic Al. Specifically, metallic aluminum serves as an effective reducing agent, converting residual Sn$^{4+}$ species into Sn$^{2+}$ through a spontaneous redox reaction, as described by:

$$3\ Sn^{4+} + 2\ Al \rightarrow 3\ Sn^{2+} + 2\ Al^{3+} \quad (1)$$

To further investigate the chemical environment, X-ray photoelectron spectroscopy (XPS) was conducted after the growth of CsSnI$_3$ nanowires (Figure 2c). The CsSnI$_3$ films grown on metallic Al exhibited a significantly higher Sn$^{2+}$/Sn$^{4+}$ ratio of 96.9%/3.1%, compared to 71.3%/28.7% on the Al$_2$O$_3$ surface, highlighting the strongly reducing conditions provided by the metallic Al interface.

When the CsSnI$_3$ precursor directly contacts the exposed Al regions, a localized reductive environment is established in which electrons are transferred from Al to Sn$^{4+}$. This process converts Al to Al$^{3+}$ while restoring the perovskite precursor to a Sn$^{2+}$-dominant state. Such interfacial redox chemistry effectively suppresses Sn$^{4+}$ formation and stabilizes the CsSnI$_3$ phase during nanowire nucleation and growth, enabling the growth of high-quality CsSnI$_3$ nanowires within the patterned regions.

The phase transition behavior during annealing was examined by X-ray diffraction (XRD). As shown in Figure S6, the unannealed films exhibit the characteristic diffraction peaks of yellow δ-CsSnI$_3$. After annealing at 160 °C, these δ-phase peaks weaken markedly, while the γ-CsSnI$_3$ (202) peak becomes dominant, confirming a clear δ→γ phase transition.

The optical properties of the material were further characterized using UV-Visible absorption spectroscopy and photoluminescence (PL) spectroscopy (Figure 2d): the absorption spectrum of the nanowire film shows a distinct absorption onset at approximately 1.26 eV, while the PL spectrum exhibits an emission peak centered at 980 nm. These two sets of results mutually corroborate, accurately determining the optical bandgap of γ-CsSnI$_3$. Additionally, benefiting from the high aspect ratio of the nanowires, the material demonstrates excellent light-trapping properties, which can extend the optical path length and further optimize light absorption efficiency—laying a solid foundation for enhancing the photoresponse performance of the device.

To optimize the energy-level alignment between CsSnI$_3$ and the NiO$_x$ HTL, 3-fluoro-2-nitroanisole (3F-2NA) was selected as the interfacial dipole modifier. This choice is motivated by its substantial molecular dipole moment of 6.85 Debye, calculated using CASTEP,[46–49] and its high *pKa* value of 34.75, estimated using MarvinSketch.[50,51] The combination of a strong intrinsic dipole and high acidity resistance enables efficient dipole formation while suppressing interfacial deprotonation, thereby mitigating perovskite degradation.[52,53]

As shown by the molecular electrostatic potential (MEP) map in Figure S7, 3F-2NA exhibits a strong dipole vector directed from the electron-rich methoxy group toward the electron-withdrawing nitro group. To further elucidate the adsorption behavior underlying for the observed interfacial modulation, density functional theory (DFT) calculations were performed for three representative configurations of 3F-2NA on the CsSnI$_3$ (202) facet (Figures 3a–3c). The calculated binding energies for the nitro-down, parallel, and aromatic-down configurations are −1.01, −0.81, and −0.63 eV, respectively.

Among these configurations, the nitro-down orientation possesses the lowest energy, indicating the most thermodynamically favorable adsorption geometry.

Ultraviolet photoelectron spectroscopy (UPS) provides direct experimental evidence for this dipole-induced interfacial energy-level modulation. The full UPS spectra in Figure S8 show clear changes in both the secondary-electron cutoff and the valence-band onset before and after 3F-2NA treatment. From these spectra, the work function (Φ) of $CsSnI_3$ is extracted to decrease from 5.51 eV to 5.40 eV upon 3F-2NA modification, accompanied by an upward shift of the valence-band maximum (VBM) from -5.82 eV to -5.66 eV. These extracted values are summarized in the band diagram presented in Figure 3d, which shows that the $CsSnI_3$ VBM becomes more closely aligned with that of $NiO_x$ (-5.40 eV).[52–54] This improved energetic alignment effectively reduces the interfacial barrier for hole transfer. Collectively, these results demonstrate that 3F-2NA forms an oriented interfacial dipole layer that shifts the vacuum level, optimizes the band alignment, and consequently facilitates more efficient hole extraction across the $CsSnI_3$/$NiO_x$ interface.

Complementary Fourier-transform infrared (FTIR) spectroscopy (Figure 3e) provides further evidence of the strong interfacial interactions induced by 3F-2NA. After surface modification, the asymmetric –$NO_2$ stretching mode exhibits a clear red-shift from 1516 $cm^{-1}$ to 1508 $cm^{-1}$. Such systematic red-shifts reflect a weakening of the corresponding vibrational bonds, which is typically associated with electronic redistribution arising from coordination or hydrogen-bond interactions between the nitro group of 3F-2NA and the surface species of $CsSnI_3$. These spectral signatures provide direct molecular-level evidence that 3F-2NA chemically anchors to the perovskite layer, thereby supporting the dipole-formation mechanism proposed in this work.

To quantitatively assess the impact of the 3F-2NA dipolar layer on carrier recombination, time-resolved photoluminescence (TRPL) measurements were conducted on $CsSnI_3$ nanowires films with and without surface modification (Figure 3f). The decay kinetics were fitted using a bi-exponential model, and the fitting parameters are summarized in Table S1. The average carrier lifetime increases from 58.7 ns (pristine) to 82.5 ns (3F-2NA treated). This lifetime prolongation suggests a possible reduction in nonradiative recombination pathways at the perovskite interface, which could be associated with improved interfacial passivation induced by the dipolar modifier.

As described in the subsequent fabrication steps shown in Figure S10, a compact $NiO_x$ layer was first deposited onto the $CsSnI_3$ nanowire film via radio-frequency (RF) magnetron sputtering to serve as the HTL. Subsequently, the transparent ITO top electrode was deposited through a shadow-mask-assisted sputtering process, followed by PMMA encapsulation to block moisture and oxygen while maintaining high optical transparency. Selected regions of the ITO electrode were intentionally left uncovered to allow direct probe access during electrical measurements, thereby minimizing contact resistance and ensuring reliable device characterization. The bottom electrode is provided by the metallic Al substrate itself, which directly contacts the $CsSnI_3$ nanowire layer through the laser-defined growth windows. Together, these steps yield the flexible $CsSnI_3$ nanowire photodetector architecture illustrated in Figure 4a.

A photograph of the encapsulated flexible device featuring 4 × 4 sensing units is shown in Figure 4b. With the device structure established, we next evaluated the electrical performance of the flexible $CsSnI_3$ photodetector under varying illumination conditions. Photocurrent–voltage (I–V) measurements were conducted from dark conditions up to an incident power density of $2.5 \times 10^2$ mW·cm$^{-2}$, and the corresponding I–V curves are presented in Figure 4c. The device exhibits a very low dark current on the order of $10^{-10}$-$10^{-9}$ A, and its open-circuit voltage increases from 0 V in the dark to approximately 0.12 V under high illumination. As the light intensity increases, the photocurrent rises monotonically, reflecting the efficient photocarrier generation and extraction efficiency of the nanowire-based architecture.

As shown in Figure 4d, the $CsSnI_3$ nanowire photodetector exhibits stable and intensity-dependent photoresponse behavior under 850 nm illumination. Each illumination level produces a distinct photocurrent step, with the amplitude decreasing systematically as the incident power density decreases from $8.39 \times 10^2$ to $1.33 \times 10^{-5}$ mW·cm$^{-2}$. The dark current remains below ~$10^{-9}$ A across all conditions, yielding an on/off ratio exceeding $10^6$. This large photocurrent modulation and well-resolved response steps highlight the device's excellent photoswitching characteristics and extremely low noise floor. The linear dynamic range (LDR) of the device was further evaluated, as shown in Figure 4e, where LDR was calculated via the formula:

$$LDR = 20 log \left(\frac{P_{sat}}{P_{low}}\right) \qquad (3)$$

where $P_{sat}$ represents the highest incident power density before the photocurrent deviates from linearity, and $P_{low}$ corresponds to the minimum detectable power density (both in mW cm$^{-2}$). The fitted photocurrent-power relationship exhibits a broad linear regime spanning several orders of magnitude, yielding an LDR of 156 dB. Such an exceptionally wide dynamic range enables the device to reliably distinguish optical signals from extremely weak to intense illumination, underscoring its strong potential for weak-light detection and high-fidelity sensing applications. To quantitatively evaluate the photodetector performance, the responsivity (R) was calculated according to:

$$R = \frac{I_{light} - I_{dark}}{P_{in} \cdot S} \qquad (2)$$

where $I_{light}$ and $I_{dark}$ are the steady-state photocurrent and dark current (A), $P_{in}$ is the incident light power density (W·cm$^{-2}$), and S is the effective area of the device (cm$^2$). Based on the obtained responsivity and dark current values, the specific detectivity *(D\*)* was derived under the shot-noise-limited assumption:

$$D^* = \frac{R}{\sqrt{2qI_{dark}/S}} \qquad (3)$$

where q stands for the elementary charge (1.602×10$^{-19}$ C), and other parameters follow the definitions used above.

As shown in Figure 4f, the responsivity decreases linearly with increasing incident power density, reaching 0.39 A W$^{-1}$ under weak illumination. Correspondingly, the specific detectivity reaches 1.38 × 10$^{13}$ Jones, benefiting from the low dark current and efficient carrier extraction. In addition, the device exhibits a −3 dB bandwidth of 2128 Hz (Figure 4g), together with fast rise/decay times of 30/30 ms (Figure 4h). These results demonstrate that the combined reductive contact engineering and dipolar interface modification effectively enhance the switching speed of the $CsSnI_3$ photodetector.

Compared with previously reported Sn- and Pb-based perovskite photodetectors (Table S2), our $CsSnI_3$ nanowire device delivers competitive or superior performance, particularly in responsivity, detectivity, and required operating bias. Its broad 156 dB linear dynamic range and low-voltage operation (0.1 V) arise from the vertically aligned nanowire architecture, the Al-induced suppression of $Sn^{4+}$ formation, and the 3F-2NA dipole–enhanced interfacial alignment. These synergistic factors collectively enable efficient NIR detection with low noise and strong operational stability, positioning this device among the most capable lead-free perovskite photodetectors reported to date.

Beyond optoelectronic performance, device stability is essential for the practical deployment. The durability of the $CsSnI_3$ nanowire photodetector was assessed in terms of continuous light operation, ambient storage stability, and mechanical flexibility.

To evaluate the device stability under prolonged illumination, the photodetector was subjected to repeated ON/OFF cycling of 850 nm laser for 6000 s. As shown in Figures 5a and 5c, the PMMA-encapsulated device maintains highly stable photocurrent, retaining over 98% of its initial value without noticeable degradation. In contrast, the unencapsulated device shows significant performance decay (Figures 5b and 5d), highlighting the essential role of PMMA encapsulation in preserving operational stability.

The device also exhibits excellent long-term environmental stability. As shown in Figure 5e, the PMMA-encapsulated photodetector retains approximately 85% of its initial photocurrent after 60 days of ambient storage, as verified through periodic ON/OFF measurements performed at different aging intervals. This long-term robustness arises from the synergistic protection of the 3F-2NA dipole layer, which mitigates interfacial degradation, and the PMMA overlayer, which effectively blocks moisture and oxygen. Together, these passivation pathways suppress $Sn^{2+}$ oxidation and ensure sustained operational durability under ambient conditions.

The mechanical durability of the device was evaluated through repeated bending tests.

As shown in Figure 5f, the photodetector maintains ~94% of its initial photocurrent after 1000 bending cycles at a fixed bending radius, with no noticeable distortion in its ON/OFF response. This high durability highlights the strong mechanical tolerance of the vertically aligned $CsSnI_3$ nanowires and the flexible Al-based substrate, ensuring reliable photoresponse under continuous mechanical deformation.

To further evaluate the device uniformity and its potential for spatial sensing, a 4 × 4 $CsSnI_3$ nanowire photodetector array was fabricated and characterized. As illustrated in Figure 5g, an H-shaped mask was placed above the array during 850 nm illumination to generate a patterned light field on the flexible device platform. The resulting spatial photocurrent maps (Figure 5h,i) reveal a clear contrast between masked and illuminated regions, demonstrating well-defined pixel-level responses across the entire array. All sixteen detector units show consistent photocurrent intensity and negligible dark-current variation, confirming the excellent uniformity, scalability, and pattern-recognition capability of the $CsSnI_3$ nanowire photodetector array.

**Conclusion**

In summary, we have developed a flexible $CsSnI_3$ nanowire photodetector through a synergistic interface-engineering strategy that combines aluminium-substrate reduction with dipolar molecular modification. The exposed Al contact provides a strong reductive environment that suppresses $Sn^{4+}$ formation during nanowire crystallization, while the oriented 3F-2NA dipole layer modulates the vacuum level, optimizes the $CsSnI_3$/$NiO_x$ band alignment, and enhances interfacial charge extraction. Benefiting from these co-engineering effects and PMMA encapsulation, the device delivers a responsivity of 0.39 A $W^{-1}$, a specific detectivity of 1.38 × $10^{13}$ Jones, and excellent operational durability—retaining 85% of its photocurrent after 60 days of ambient storage and 94% after 1000 bending cycles.

These results establish a clear structure–property relationship linking reductive contact chemistry and dipole-induced energy-level tuning to improved stability and performance in tin-based perovskite optoelectronics. The insights and design principles demonstrated here offer a generalizable pathway for advancing high-performance, lead-free, and flexible perovskite photodetectors and related low-dimensional Sn-based optoelectronic systems.

**Data Availability Statement**

The data that support the findings of this study are available from the corresponding author upon reasonable request.

**Acknowledgements**

This work was financially supported by the Hubei Provincial Natural Science Foundation of China [Grant No. 2023AFB353] and the Innovation Project of Optics Valley Laboratory [Grant No. OVL2023PY004]. The authors sincerely thank Prof. Mingkui Wang at Huazhong University of Science and Technology for his insightful guidance throughout this work.

**Author contributions**

L.D. conceived and designed the project. L.D. fabricated the flexible $CsSnI_3$ nanowire photodetectors and carried out all device fabrication, electrical measurements, optical characterizations, and stability tests. W.C. and Q.G. assisted with device fabrication and electrical characterization. Y.X. contributed to material synthesis optimization and optical measurements. G.Z. provided technical support on laser patterning and aluminum substrate processing. N.C. performed DFT calculations and assisted with theoretical analysis of the dipolar interface modification. L.D. conducted data analysis and prepared all figures. X.L. provided critical discussions, supervised part of the experimental design, and contributed to data interpretation. L.D. wrote the manuscript with input from all authors. All authors reviewed and approved the final manuscript.

**Conflict of Interest**

The authors declare no conflict of interest.

**Additional information**

**Supplementary information**

The online version contains supplementary material available at

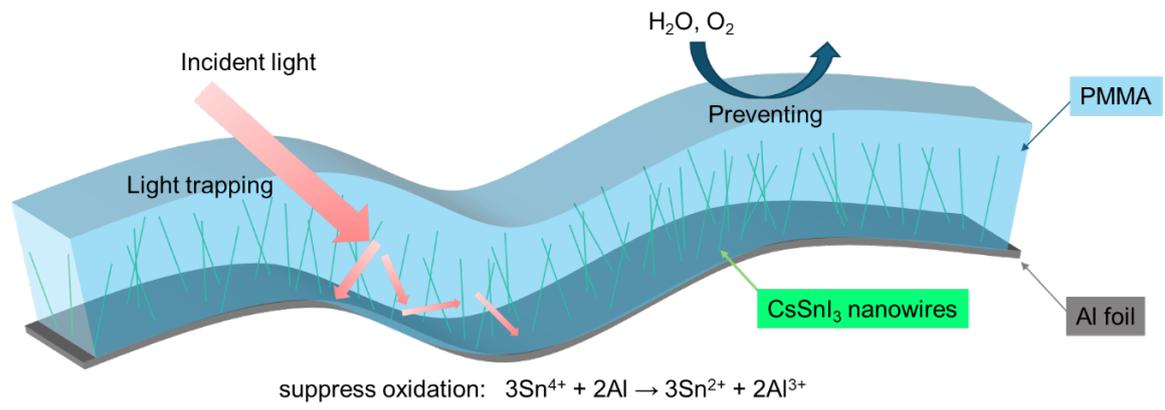

Figure 1. Schematic illustration of the flexible CsSnI$_3$ nanowire photodetector design.

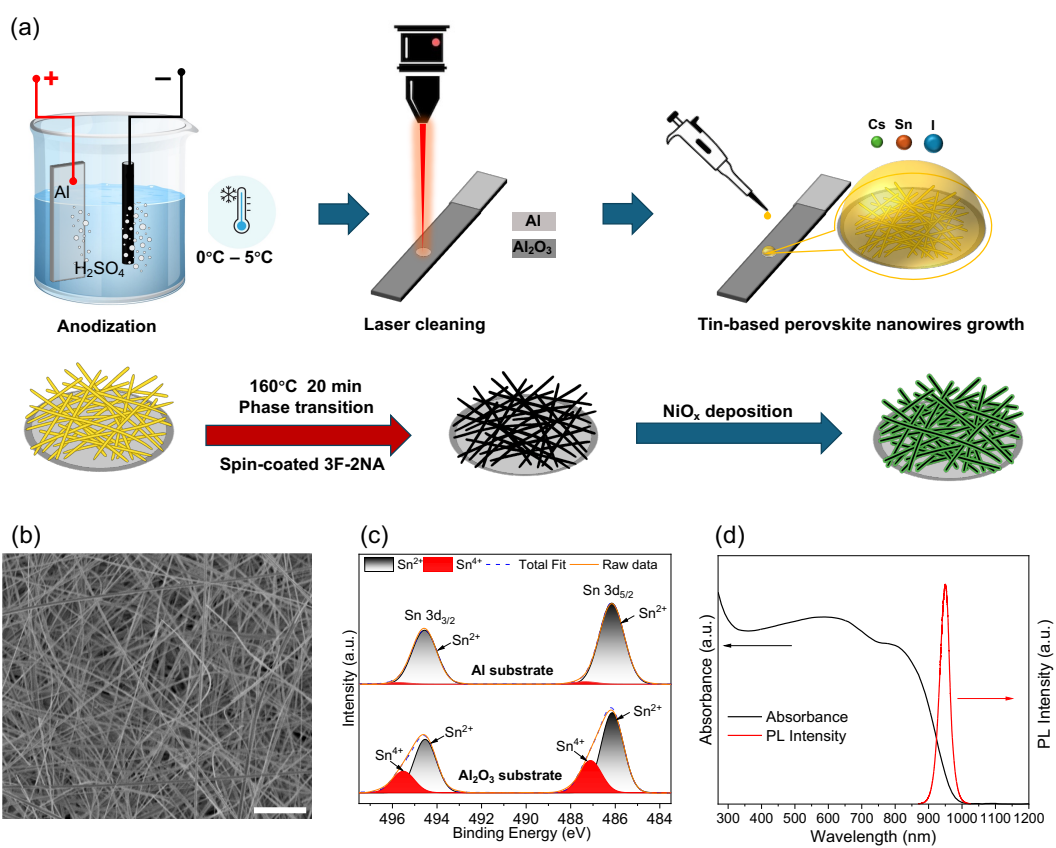

Figure 2. (a) Schematic representation of the partial fabrication steps of the CsSnI$_3$ nanowire photodetector; (b) SEM image of vertically aligned CsSnI$_3$ nanowires grown on metallic Al, scale bar: 5 μm; (c) Sn 3d XPS spectra of CsSnI$_3$ films grown for 12 h on metallic Al and anodized Al$_2$O$_3$; (d) UV–vis absorption and PL spectra of CsSnI$_3$ nanowires.

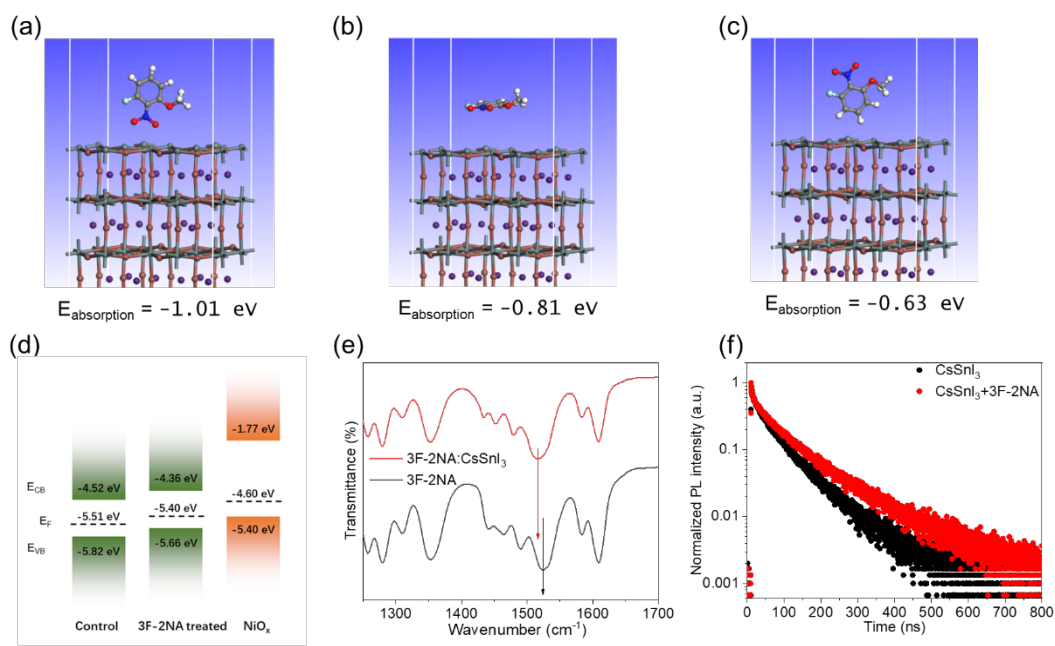

Figure 3. (a–c) DFT-optimized adsorption configurations of 3F-2NA on the CsSnI$_3$ (202) facet; (d) Energy-level diagram of CsSnI$_3$ before and after 3F-2NA modification, derived from the UPS spectra shown in Figure S8, together with the corresponding band alignment relative to NiO$_x$; (e) FTIR spectra of CsSnI$_3$ with and without 3F-2NA treatment; (f) TRPL decay curves of CsSnI$_3$ nanowires with and without 3F-2NA modification.

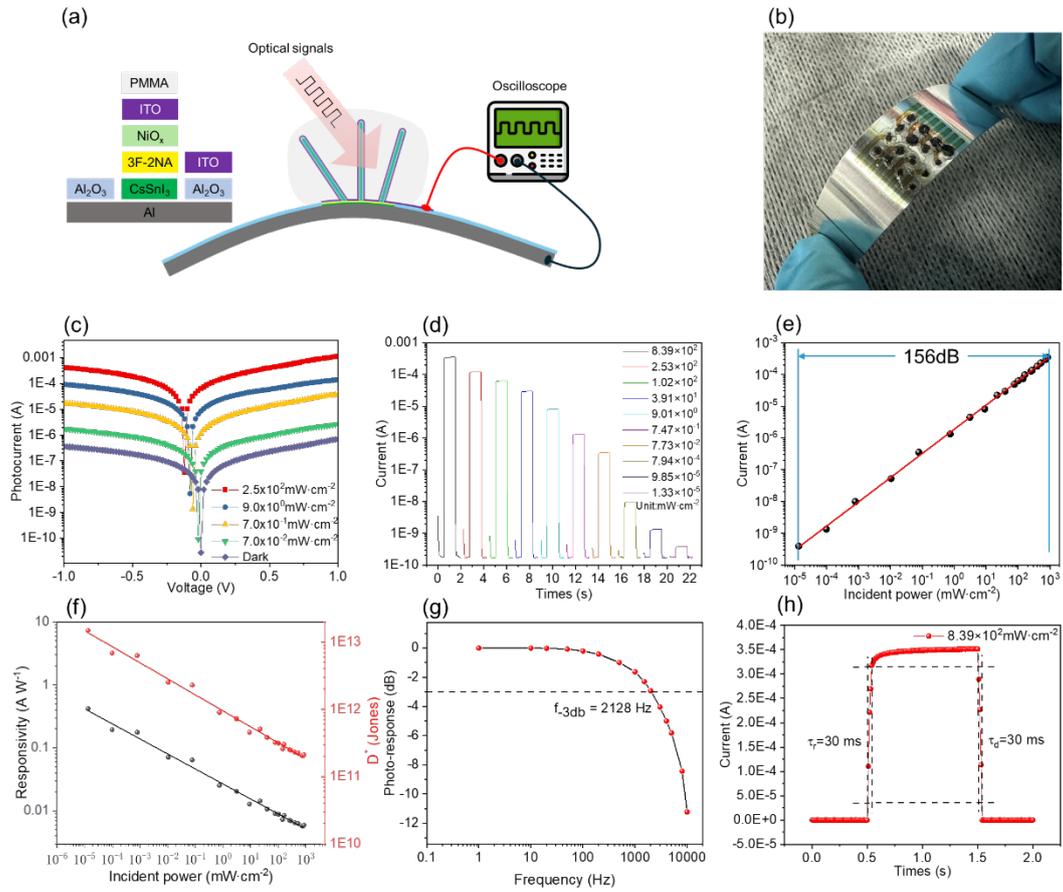

Figure 4. (a) Cross-sectional device structure and operating principle of the $CsSnI_3$ nanowire photodetector; (b) Photograph of the flexible $CsSnI_3$ nanowire photodetector; (c) I–V characteristics measured under different illumination intensities and in the dark; (d) Time-resolved photoresponse recorded at various incident power densities; (e) Linear dynamic range (LDR) of the device; (f) Responsivity and detectivity as functions of incident optical power; (g) −3 dB cutoff frequency of the device; (h) Rise and decay times measured under an illumination intensity of $8.39 \times 10^2$ mW·cm$^{-2}$.

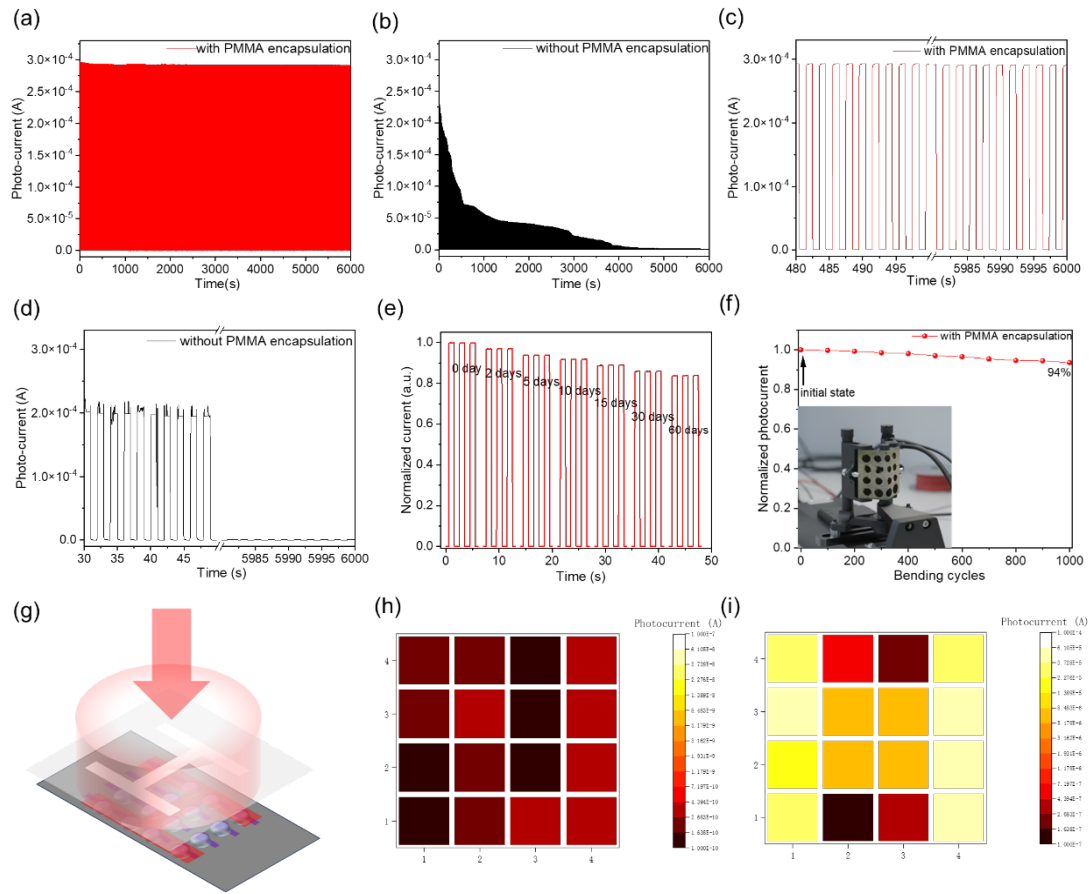

Figure 5. (a,b) Photocurrent stability of the unencapsulated and PMMA-encapsulated CsSnI$_3$ nanowire photodetectors under continuous 850 nm illumination modulated at 1 Hz over 6000 s; (c,d) Enlarged views of the corresponding stability traces; (e) Photocurrent retention of the PMMA-encapsulated device after 60 days of ambient storage; (f) Normalized photocurrent evolution over 1000 bending cycles; (g) Schematic illustration of the 4 × 4 photodetector array and a representative device under partial masking during 850 nm illumination. (h,i) Spatial photocurrent maps of the array measured in the dark and under illumination.

# Supporting Information

**Reductive Contact and Dipolar Interface Engineering Enable Stable Flexible CsSnI$_3$ Nanowire Photodetectors**

*Letian Dai[a,b,*], Wanru Chen[c], Quanming Geng[a], Ying Xu[a], Guowu Zhou[d], Nuo Chen[e], Xiongjie Li[f,*]*

[a] Wuhan National Laboratory for Optoelectronics, Huazhong University of Science and Technology, 1037 Luoyu Road, Wuhan 430074, Hubei, P. R. China

[b] Optics Valley Laboratory, Hubei 430074, P. R. China

[c] Wuhan Huaming Renewables Co., Ltd., 122 Zhongbei Road, Wuhan 430077, Hubei, P.R. China

[d] Wuhan Haisipulin Technology Development Co., Ltd, 897 Gexin Road, Wuhan, 430023, Hubei, P.R. China

[e] Department of Electrical Engineering, Karlsruhe Institute of Technology, Karlsruhe 76131, Germany

[f] Key Laboratory of Catalysis and Energy Materials Chemistry of Ministry of Education & Hubei Key Laboratory of Catalysis and Materials Science, Key Laboratory of Analytical Chemistry of the State Ethnic Affairs Commission, Hubei R&D Center of Hyperbranched Polymers Synthesis and Applications, South-Central Minzu University, Wuhan, Hubei, 430074, P.R. China

*Email: ldai@hust.edu.cn, xiongjieli@mail.scuec.edu.cn.

## S1. Materials and Preparation

CsI (99.99%, Alfa Aesar), SnI$_2$ (99.99%, Sigma-Aldrich), DMF (*N,N*-dimethylformamide, 99.8%, Sigma-Aldrich), DMSO (Dimethyl sulfoxide, 99.5%, Sigma-Aldrich), PMMA (Poly(methyl methacrylate), average Mw ~15,000 by GPC, Macklin), Cabon rod (100mm, diameter 5.0mm, Macklin) and 3F-2NA (3-Fluoro-2-nitroanisole, 99.7%, Macklin) were used as received without further purification. Aluminum foil (type 1060) with a thickness of 0.2 mm and a high purity ≥ 99.60% was obtained from an enterprise in Hubei province of China. A 200 mL aqueous solution of 18 wt% sulfuric acid was prepared by slowly adding 22 mL of concentrated H$_2$SO$_4$ (98%, Macklin) into 178 mL of deionized water under constant stirring and cooling.

## S2. Fabrication of CsSnI$_3$ Nanowire Photodetector

Prior to use, the aluminum foil was ultrasonically cleaned in ethanol for 15 min and dried under nitrogen. Anodization was performed in 18 wt% H$_2$SO$_4$ at 0–5 °C under 25 V for 60 min to form an anodic Al$_2$O$_3$ layer. After rinsing and drying, a nanosecond 650 nm pulsed laser was used to locally ablate the Al$_2$O$_3$ layer, exposing the metallic Al regions that act as the conductive base.

In a glovebox environment with O$_2$ and H$_2$O levels maintained below 0.01 ppm, a 1.5 M CsSnI$_3$ precursor solution (prepared by dissolving CsI and SnI$_2$ in a stoichiometric ratio of 1:1 in DMF) was drop-cast onto the laser-exposed Al surface and dried at room temperature for 12 h under a controlled solvent evaporation rate of ~1 μL h$^{-1}$. During this slow-drying process, γ-CsSnI$_3$ nanowires spontaneously formed. The resulting nanowire film was subsequently annealed at 160 °C for 20 min, leading to the formation of the black δ-CsSnI$_3$ phase. After phase transformation, a 5 wt% 3F-2NA solution in chlorobenzene was spin-coated onto the nanowire array at 3000 rpm for 30 s, followed by a mild post-annealing step at 80 °C for 10 min to form a stable interfacial dipole layer.

Subsequently, a compact NiO$_x$ layer (20 nm) serving as the hole transport layer (HTL), followed by a 250 nm ITO top transparent electrode, were sequentially deposited onto the CsSnI$_3$ nanowires by sputtering.

Finally, the device was encapsulated with a thin PMMA layer to prevent the oxidation of Sn²⁺ species.

## S3. Characterization Techniques

XRD measurements were carried out using a PANalytical X'Pert PRO diffractometer (Cu Kα, λ = 1.5406 Å). XPS and UPS spectra were collected using a Shimadzu AXIS-ULTRA DLD-600 W system. Surface morphology was examined using a Nova Nano SEM 450. UV–Vis absorption spectra were recorded using a Lambda 950 spectrophotometer, while steady-state PL and TRPL spectra were measured using an Edinburgh FLS 980 system with a 405 nm excitation source.

The optoelectronic properties of the flexible $CsSnI_3$ nanowire photodetector were evaluated at room temperature under 850 nm laser illumination. A focused beam (~3 mm diameter) illuminated the active area (0.07 cm²), with the power density adjustable from $1.33 \times 10^{-5}$ to $8.39 \times 10^{2}$ mW·cm⁻². Photoresponse curves were recorded with a Keithley 2400 source meter. Under a 0.1 V bias and 850 nm illumination, the power-dependent photocurrent, linear dynamic range, ON/OFF ratio, and temporal response were measured. The photocurrent shows linear dependence over the range of $1.33 \times 10^{-5}$–$8.39 \times 10^{2}$ mW·cm⁻².

## S4. Time-Resolved Photoluminescence (TRPL) Analysis

The time-resolved photoluminescence (TRPL) decay curves of $CsSnI_3$ nanowire films were fitted using a bi-exponential function, as shown in Equation (S1):

$$I(t) = A_1 e^{-\frac{t}{\tau_1}} + A_2 e^{-\frac{t}{\tau_2}} + I_0 \tag{S1}$$

where $A_1$ and $A_2$ are the relative amplitudes, $\tau_1$ and $\tau_2$ are the fast and slow decay lifetimes, and $I_0$ is a baseline offset.

The average carrier lifetime $\tau_{average}$ was calculated using Equation (S2):

$$\tau_{average} = \frac{A_1\tau_1^2 + A_2\tau_2^2}{A_1\tau_1 + A_2\tau_2} \tag{S2}$$

The fitted parameters ($A_1, A_2, \tau_1, \tau_2$) were obtained from the bi-exponential fitting of Figure S9.

**S5. Supplementary Figures and Tables**

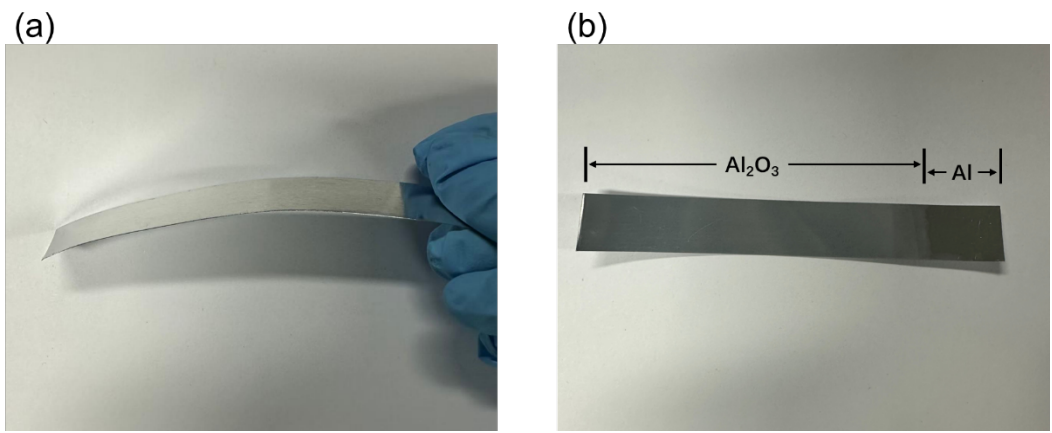

Figure S1. Photographs of 0.2 mm-thick 1060 aluminum foil before and after anodization. (a) Pristine smooth and flexible foil; (b) surface covered by a gray-white $Al_2O_3$ layer after anodization.

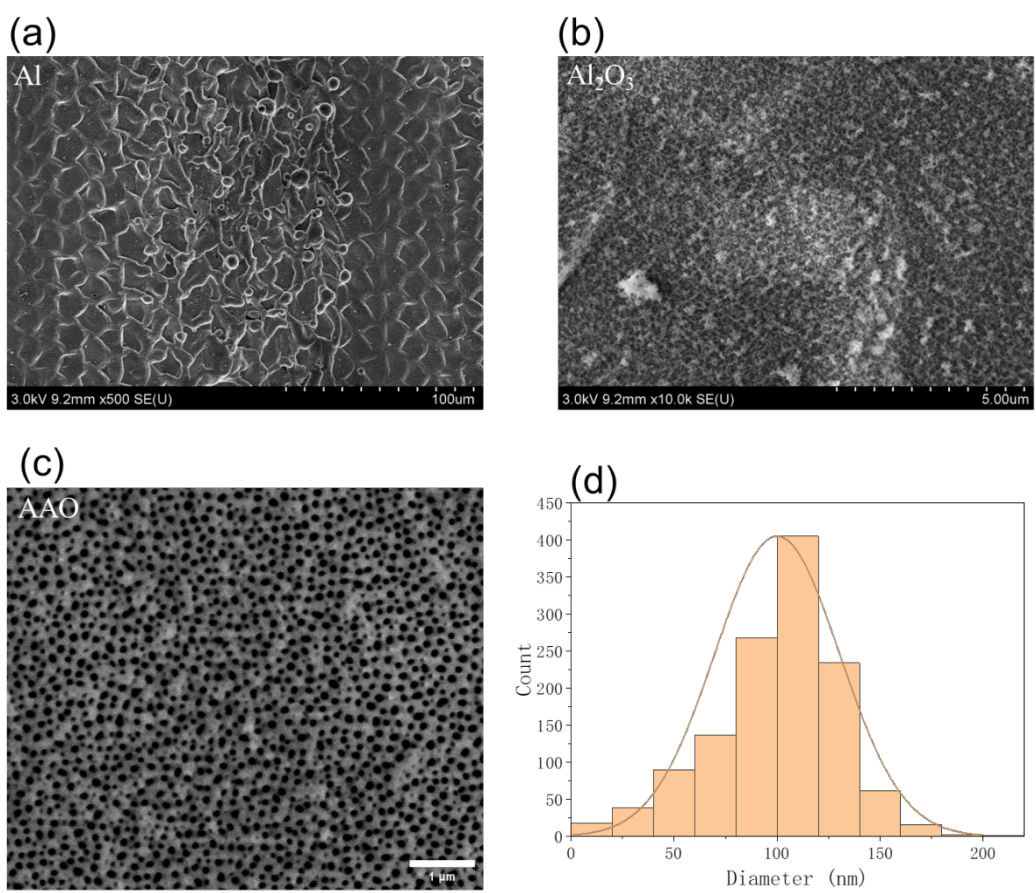

Figure S2. Surface morphology of aluminum foil before and after anodization. (a, b) scanning electron microscopy (SEM) images of pristine and anodized Al foil; (c) Top-view SEM image of anodic aluminum oxide (AAO) layer with ~100 nm pores (Scale bar: 1 μm); (d) Corresponding pore size distribution extracted from the SEM image in (c).

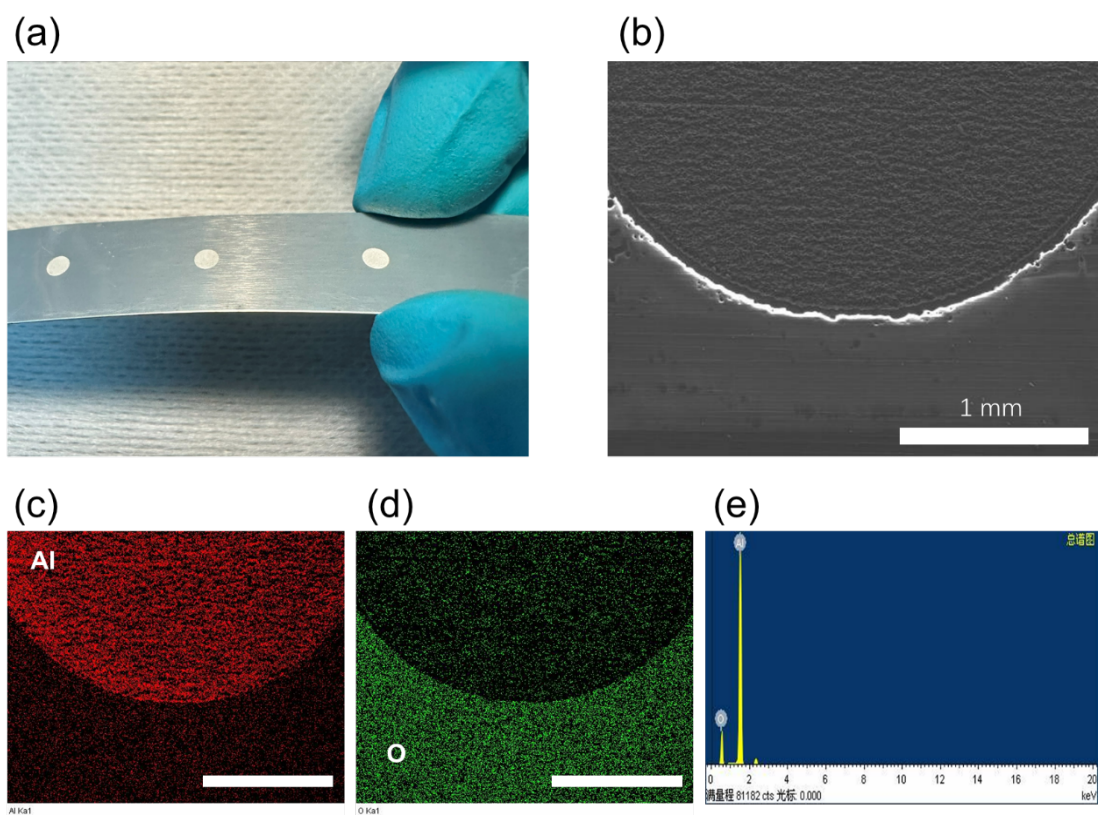

Figure S3. (a) Photograph of circular laser-ablation windows on the anodized Al foil; (b) SEM image of a localized laser-ablated region, showing the clear boundary between the exposed metallic Al and the remaining $Al_2O_3$ layer (scale bar: 1 mm); (c, d) EDX elemental maps of Al and O, respectively (scale bars: 1 mm); (e) Corresponding EDX spectrum of anodized Al foil.

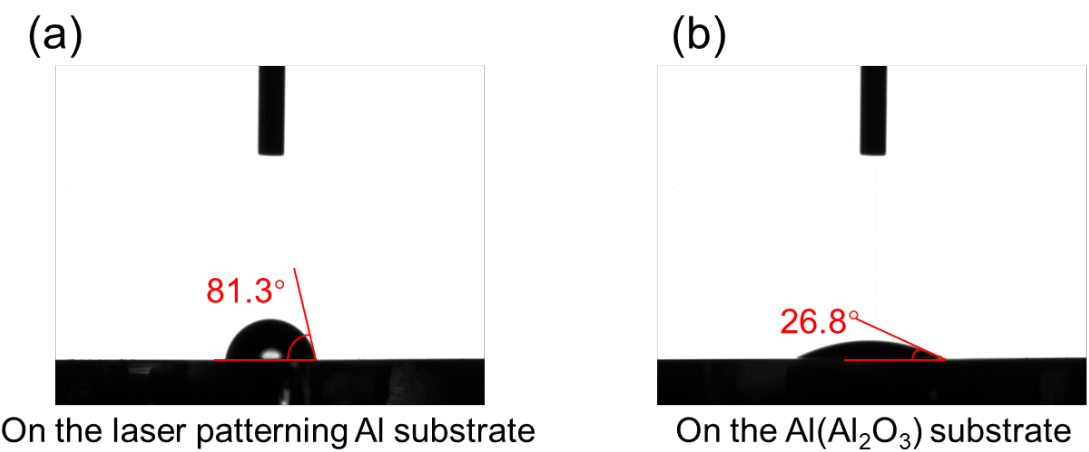

Figure S4. Contact-angle measurements of CsSnI$_3$ precursor solution. (a) On laser-ablated Al surface (81.3°); (b) On anodized Al$_2$O$_3$ surface (26.8°).

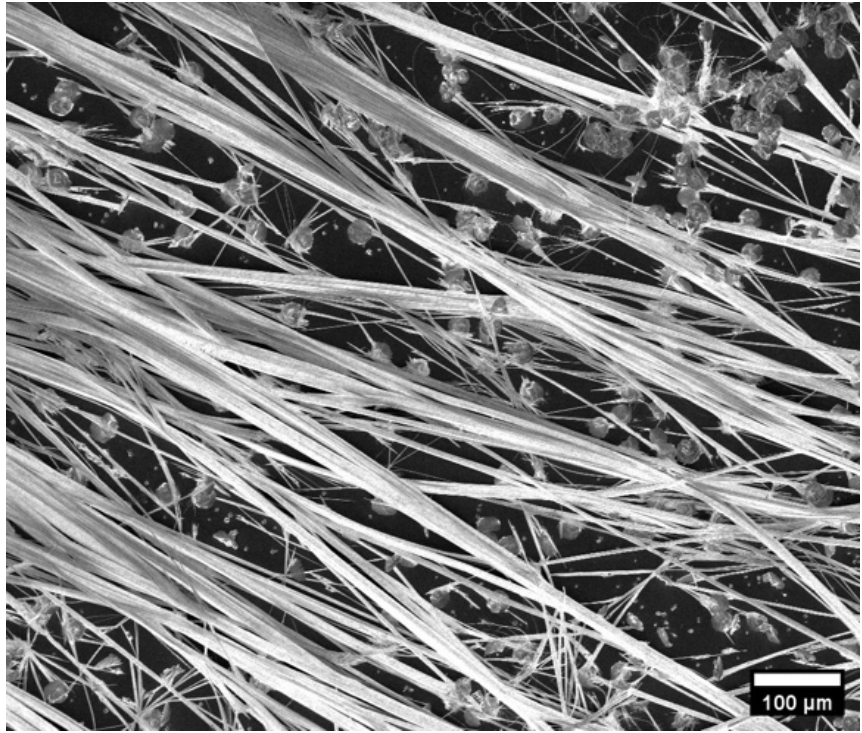

Figure S5. SEM image showing the lateral growth morphology of $CsSnI_3$ microwires formed on the anodized $Al_2O_3$ surface (scale bar: 100 μm).

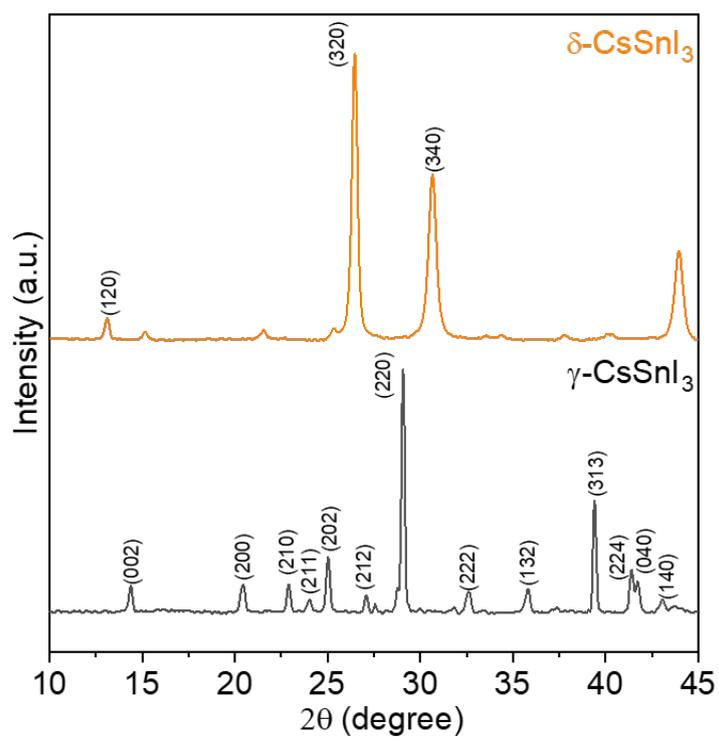

Figure S6. X-ray diffraction (XRD) patterns of CsSnI$_3$ showing the yellow δ-phase before annealing and the black γ-phase after annealing at 160 °C

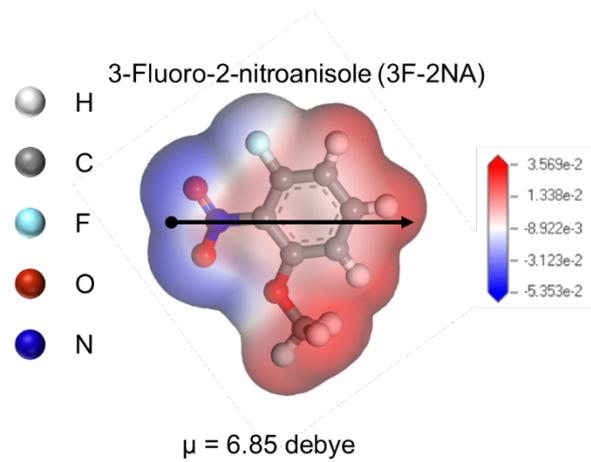

Figure S7. Molecular electrostatic potential (MEP) map of 3F-2NA.

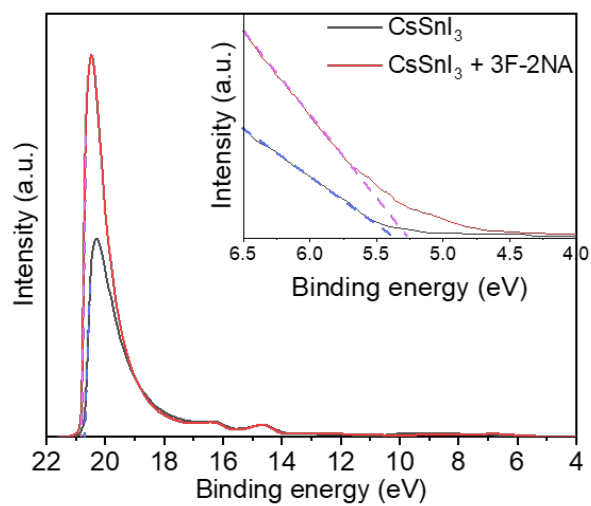

Figure S8. Ultraviolet photoelectron spectroscopy (UPS) spectra of pristine and 3F-2NA-modified $CsSnI_3$ films.

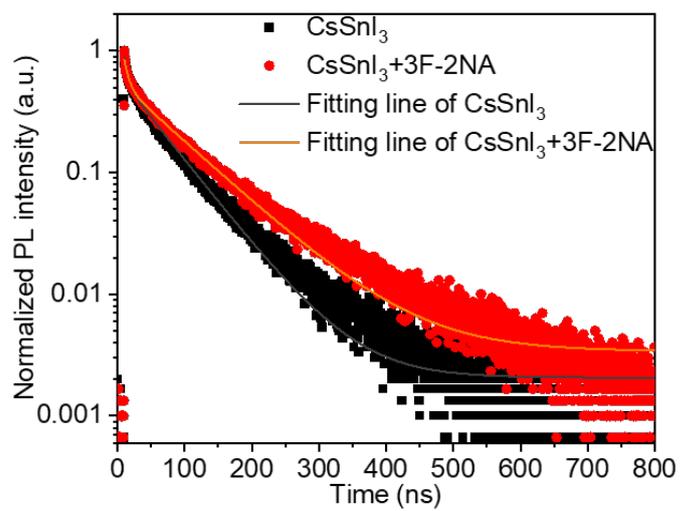

Figure S9. TRPL decay curves of CsSnI$_3$ nanowires with and without 3F-2NA modification, with bi-exponential fitting.

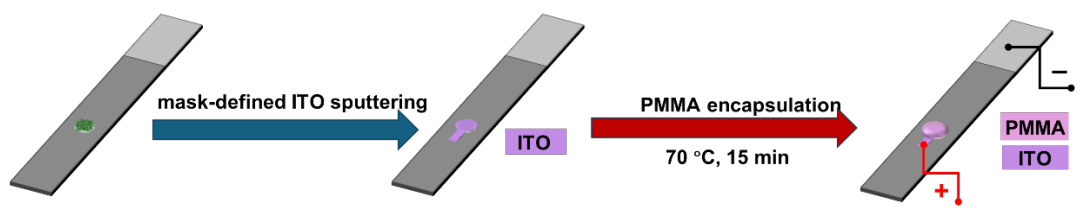

Figure S10. Shadow-mask-assisted deposition of ITO followed by PMMA encapsulation.

Table S1 Bi-exponential TRPL fitting parameters for pristine and 3F-2NA-modified CsSnI$_3$.

|  | A$_1$ | τ$_1$ (ns) | A$_2$ | τ$_2$ (ns) | y$_0$ | t$_{avg}$ (ns) |
|---|---|---|---|---|---|---|
| CsSnI$_3$ | 0.331 | 6.334 | 0.537 | 61.963 | 0.002 | 58.67 |
| CsSnI$_3$ + 3F-2NA | 0.343 | 6.714 | 0.503 | 86.487 | 0.003 | 82.47 |

Table S2 Performance comparison between this work and previously reported perovskite photodetectors.

| Perovskites | Wavelength (nm) | Responsivity (mA/W) | Detectivity (Jones) | Response time rise/decay time | LDR (dB) | Bias (V) | refs |
|---|---|---|---|---|---|---|---|
| $CsSnI_3$ NW array | 850 | 390 | $1.38\times10^{13}$ | 30 ms/30 ms | 156 | 0.1 | This work |
| $CsSnI_3$ NW array | 940 | 54 | $3.85\times10^{5}$ | 83.8 ms/243.4 ms | | 0.1 | [1] |
| $CsSnI_3$ NW array | 405 | 237 | $1.18\times10^{12}$ | 230 μs/190 μs | 180 | | [2] |
| $CsPbI_3$ NW array | 630 | $1.294\times10^{6}$ | $2.6\times10^{14}$ | 0.85 ms/0.78 ms | | 5 | [3] |
| $CsPbI_3$ NW array | 530 | 350 | $1.64\times10^{12}$ | | | 1 | [4] |
| $CsPbI_3$ NW array | Halogen light | 6.7 | $1.58\times10^{8}$ | 292μs/234 μs | | 5 | [5] |
| $CsSnBr_3$ NSs | 442 | 640 | | 19 μs/ 24 μs | | 5 | [6] |
| $CsPb_xSn_{1-x}Br_3$ NWs | 473 | 110 | $2\times10^{10}$ | 4.25ms/4.82ms | 120 | 1 | [7] |

Note: NW is nanowire, NR is nanorod, MW is microwire, NP is nanoplatelet, NS is nanosheet, TF is thin film, MC is microcrystal, MR is microrod.